\documentclass[10pt]{ijnam}
\def\currentvolume{1}
\def\currentissue{1}
\def\currentyear{2008}
\def\month{January}
\pagespan{1}{10}
\copyrightinfo{2008}{} % copyright year

\def\dr{\hbox{d}r}
\def\dt{\hbox{d}t}
\def\dw{\hbox{d}w}
\def\dd{\hbox{d}}

\newtheorem{theorem}{Theorem}
\newtheorem{corollary}{Corollary}
\newtheorem{remark}{Remark}
\newtheorem{definition}{Definition}

\usepackage{natbib}

\usepackage{graphicx}
\usepackage{amssymb}

\begin{document}

\title{Approximate formulae for pricing zero-coupon bonds and their asymptotic analysis}

\author{B. Stehl\'\i kov\'a \and  D. \v Sev\v covi\v c}
\address{
 Department of Applied Mathematics and Statistics,
 Comenius University, Mlynsk\' a dolina,
 842 48 Bratislava, Slovakia
}

\email{\{stehlikova,sevcovic\}@fmph.uniba.sk}
\urladdr{http://www.iam.fmph.uniba.sk/institute/sevcovic/}

% communicated
\commby{Lubin G. Vulkov}

% Received by the editors ?
\date{January 31, 2008 and, in revised form, May 22, 2008.}


\abstract{
We analyze analytic approximation formulae for pricing zero-coupon bonds 
in the case when the short-term interest rate is driven by a 
one-factor mean-reverting process with a volatility nonlinearly depending on
the interest rate itself.  We derive the order of accuracy of the analytical approximation due to Choi and Wirjanto.
We furthemore give an explicit formula for a higher order
approximation and we test both approximations numerically for 
a class of one-factor interest rate models.
}


\maketitle
\pagestyle{myheadings}
\markboth{B. Stehl\'\i kov\'a \and  D. \v Sev\v covi\v c}{Approximate formulae for pricing zero-coupon bonds}

\section{Introduction}
\label{sec:intro}

Term structure models give the dependence of time to maturity of a discount bond and its present price. One-factor models are often formulated in terms of a stochastic differential equation for the instantaneous interest rate (short rate). In the theory of nonarbitrage term structure models the bond prices (yielding the interest rates) are given by a solution to a parabolic partial differential equation. The stochastic differential equation for the short rate is specified either under a real (observed) probability measure or risk-neutral one. A risk-neutral measure is an equivalent measure such that the derivative prices (bond prices in particular) can be computed as expected values. If the short rate process is considered with a real  probability measure, a function $\lambda$ describing the so-called  market price of risk has to be provided. The volatility part of the process is the same for both real and risk-neutral specification of the process. The changes in the drift term depend on the so called market price of risk function $\lambda$.

It is often assumed that the short rate evolves according to the following mean reverting  stochastic differential equation
\begin{equation}
\dr=(\alpha+\beta r)\dt+\sigma r^{\gamma} \dw
\label{1f-short-rate}
\end{equation}
where $\sigma>0$, $\gamma \geq 0$, $\alpha>0$, $\beta$ are given parameters. In particular, it includes the well known Vasicek model ($\gamma=0$) and Cox-Ingersoll-Ross  model ($\gamma=1/2$) 
(c.f. Vasicek \cite{V} and Cox, Ingersoll and Ross \cite{CIR}). For those particular choices of $\gamma$ closed form solutions of the bond pricing PDE (\ref{PDE-1f}) are known.
Assuming a suitable form of the market price of risk it turns out that both the real and risk neutral processes for the short rate have the form (\ref{1f-short-rate}). More details concerning the term structure modeling can be 
found in Kwok \cite{K}.

Using US Treasury Bills data (June 1964 - December 1989), the real probability model (\ref{1f-short-rate}) and generalized method 
of moments Chan {\it et al.} \cite{CKLS} estimated the parameter $\gamma$ at the value $1.499$. This is considered to be an important contribution, as it drew attention to a more realistic form of the short rate volatility (compared to Vasicek or CIR models). Using the same US Treasury Bills data,
Nowman in \cite{N} estimated  $\gamma=1.361$ by means of Gaussian methodology. It should be noted that these estimations of $\gamma$ are beyond values $\gamma=0$ or $\gamma=\frac{1}{2}$ for which the closed form solution of the bond prices is known in an explicit form. 
In \cite{TG} a model with interest rates from eight countries using 
generalized method of moments and quasi maximum likelihood method 
has been estimated. They tested the restrictions imposed by Vasicek and CIR models using the J-statistics in the generalized method of moments and likelihood ratio statistics in the quasi maximum likelihood method. In all tested cases except of one, the restrictions $\gamma=0$ or $\gamma=\frac{1}{2}$ were rejected.
Hence, the study of the bond prices for values of $\gamma$ different from $0$ and $1/2$  can be justified by empirical results. However, in these cases no closed form expression for bond prices is known. An approximate analytical solution was suggested in  
\cite{1f-approximation} which could make the models with general $\gamma >0$ to be more widely used. In this paper, we analyze the analytical approximation by 
Choi and Wirjanto \cite{1f-approximation} and derive its accuracy order. Furthemore, by adding extra terms to it we derive an improved, higher order approximation of the bond prices.

The paper is organized as follows. In the second section, we derive the order of approximation of the analytical approximative solution from 
\cite{1f-approximation}. We derive a new, higher order accurate approximation. In the third section, we compare the two approximations with a known closed form solution from the CIR model ($\gamma=\frac{1}{2}$). In Appendix we provide a proof of uniqueness of a solution of a  partial differential equation for bond pricing for the parameter range $\frac{1}{2} \leq \gamma < \frac{3}{2}$.

\section{Accuracy of the analytic approximation formula for the bond price in the one-factor interest rate model}
\label{sec:accuracy}

In \cite{1f-approximation} the authors proposed an approximate analytical formula for the bond price in a one-factor interest rate model. They considered a  model  having  a form
(\ref{1f-short-rate})
under the risk-neutral measure. It corresponds to the real measure process:
\[
\dr= \left( \alpha+\beta r+\lambda(t,r) \sigma r^{\gamma} \right) \dt+\sigma r^{\gamma} \dw
\]
where $\lambda(t,r)$ is the so called market price of risk. For a general market price of risk function  $\lambda(t,r)$, the price $P$ of a zero-coupon bond  can be obtained from a solution to the following partial differential equation:
\begin{equation}
-\partial_{\tau} P + \frac{1}{2} \sigma^2 r^{2 \gamma} \partial^2_{r} P + (\alpha + \beta r)\partial_{r} P - rP =0, \; r>0, \; \tau \in (0,T)
\label{PDE-1f}
\end{equation}
satisfying the initial condition $P(0,r)=1$ for all $r>0$ (see e.g. 
\cite[Chapter 7]{K}).

\begin{definition}
\label{def:1}
By a complete solution to (\ref{PDE-1f}) we mean a function $P=P(\tau,r)$ having continuous  partial derivatives $\partial_{\tau} P$, $ \partial_{r} P$, $ \partial^2_{r} P$ on
$Q_T = [0,\infty) \times (0,T)$, satisfying equation (\ref{PDE-1f}) on $Q_T$,  the initial condition for $r\in[0,\infty)$ and fulfilling the following growth conditions:
$\left| P(\tau,r) \right| \leq M e^{-m r^{\delta}}$ and $|P_r (\tau,r)|\le M$ for any $r>0, t\in(0,T)$, where
$M, m, \delta>0$ are constants.
\end{definition}
It is worth to note that comparison of approximate and exact solutions is meaningful only if the uniqueness of the exact solution is guaranteed. The next theorem gives us the uniqueness of a solution to (\ref{PDE-1f}) satisfying Definition \ref{def:1}. In order not to  interrupt the discussion on approximate formulae for a solution to (\ref{PDE-1f}) a PDE based proof of the uniqueness of the exact solution is postponed to Appendix.

\begin{theorem}
Assume $\frac{1}{2} < \gamma < \frac{3}{2}$ or $\gamma=\frac{1}{2}$ and $2 \alpha \geq \sigma^2$. Then there exists a unique complete solution to (\ref{PDE-1f}).
\end{theorem}

Now let us state the main result on approximation of a solution to (\ref{PDE-1f}) due to 
Choi and Wirjanto \cite{1f-approximation}. They proposed the following approximation $P^{ap}$ for the exact solution $P^{ex}$:
\begin{theorem} \cite[Theorem 2]{1f-approximation} 
The approximate analytical solution $P^{ap}$ is given by
\begin{eqnarray}
\label{1f-approximation-formula}
\ln P^{ap}(\tau,r) &=& -rB+\frac{\alpha}{\beta} (\tau-B)+ \left( r^{2 \gamma} +q \tau \right) \frac{\sigma^2}{4 \beta} \left[ B^2 + \frac{2}{\beta} (\tau-B) \right] 
\nonumber \\
& & -q \frac{\sigma^2}{8 \beta^2} \left[B^2(2 \beta \tau-1) - 2B \left(2 \tau - \frac{3}{\beta} \right) + 2 \tau^2 - \frac{6 \tau}{\beta} \right]
\end{eqnarray}
where
$q(r) = \gamma(2 \gamma -1)\sigma^2 r^{2(2 \gamma-1)} + 2 \gamma r^{2 \gamma-1} (\alpha+\beta r)$ and
$B(\tau) = (e^{\beta \tau}-1)/\beta$.
\end{theorem}

Derivation of the formula (\ref{1f-approximation-formula})  is based on calculating the price as an expected value under a risk neutral measure. The tree property of conditional expectation was used and the integral appearing in the exact price was approximated to obtain a closed form approximation.

Authors furthermore showed that such an approximation coincides with the exact solution in the case of the Vasicek model. Moreover, they compared the above approximation with the exact solution of the CIR model which is also known in a closed form (c.f. \cite{CIR}). Graphical and tabular description of the relative error in the bond prices has been also provided in \cite{1f-approximation}.

The main purpose of this paper is to derive the order of accuracy of the approximation formula (\ref{1f-approximation-formula}) by estimating the difference $\ln P^{ap} - \ln P^{ex}$ of logarithms of approximative and exact solutions of the bond valuation equation (\ref{PDE-1f}). Then, we give an approximation formula of higher order and we analyze its order of convergence analytically and numerically.

\subsection{Error estimates for the approximate analytical solution}
\label{error}
In this part we derive the order of accuracy for the approximation derived by 
Choi and Wirjanto \cite{1f-approximation}.
% ------------- theorem --- label{theorem-accuracy-for-ln} ----------
\begin{theorem}\label{theorem-accuracy-for-ln}
Let $P^{ap}$ be the approximative solution given by (\ref{1f-approximation-formula}) and $P^{ex}$ be the exact bond price given as a unique complete solution to (\ref{PDE-1f}). Then
\[
\ln P^{ap}(\tau,r) - \ln P^{ex}(\tau,r) = c_5(r) \tau^5 + o(\tau^5) 
\]
as $\tau \rightarrow 0^+$ where
\begin{eqnarray}
\label{c5(r)}
c_5(r) &=& -\frac{1}{120} \gamma r^{2(\gamma-2)} \sigma^2 
\left[ 
2 \alpha^2 (-1+2 \gamma) r^2 + 4 \beta^2 \gamma r^4 
- 8 r^{3+2\gamma} \sigma^2
\right.
\nonumber \\ 
& &
+ 2 \beta (1-5 \gamma + 6 \gamma^2) r^{2 (1+\gamma)} \sigma^2 
+\sigma^4 r^{4 \gamma} (2\gamma-1)^2 (4 \gamma-3) 
 \\
& &
\left.
+ 2 \alpha r \left( \beta(-1+4 \gamma)r^2 + (2 \gamma -1)(3\gamma -2) r^{2 \gamma} \sigma^2 \right)
 \right]. \nonumber
\end{eqnarray}
The convergence is uniform w. r. to $r$ on compact subintervals $[r_1,r_2]  \subset\subset (0,\infty)$.
\end{theorem}
%------------------------------------------------------------

%--------------------remark c_5----------------------------------------
\begin{remark}
The function $c_5(r)$ remains bounded as $r \rightarrow 0^+$ for the case of the CIR model in which $\gamma=1/2$. More precisely, $\lim_{r \rightarrow 0} c_5(r) = -\frac{\sigma^2}{120} \alpha \beta$. If  $1/2 < \gamma < 1$, then $c_5(r)$ becomes singular, 
$c_5(r)=O \left( r^{2(\gamma-1)} \right)$ as $r \rightarrow 0^+$.
\end{remark}

% -------proof of the theorem -------------
{\em Proof:} 
Recall that the exact bond price $P^{ex}(\tau,r)$ for the model (\ref{1f-short-rate}) is given by a solution of the PDE (\ref{PDE-1f}). 
Let us define the following auxiliary function: 
$
f^{ex}(\tau,r)=\ln P^{ex}(\tau,r)\,.
$
Clearly, 
$\partial_{\tau} P^{ex} = P^{ex} \partial_{\tau} f^{ex}, \partial_{r} P^{ex} = P^{ex}  \partial_{r} f^{ex}$
and\hfill \\
$\partial^2_{r} P^{ex} = P^{ex} \left[ \left( \partial_{r} f^{ex} \right)^2 +  \partial^2_{r} f^{ex} \right]$.
Hence the PDE for the function $f^{ex}$ reads as follows:
\begin{equation}
-\partial_{\tau} f^{ex} + \frac{1}{2} \sigma^2 r^{2 \gamma} \left[ \left( \partial_{r} f^{ex} \right)^2 +  \partial^2_{r} f^{ex} \right] + (\alpha + \beta r)
 \partial_{r} f^{ex} - r =0.
\label{eq-lnP}
\end{equation}
Substitution of $f^{ap} = \ln P^{ap}$ into equation (\ref{eq-lnP}) yields a nontrivial right-hand side $h(\tau,r)$ for the equation for the approximative solution $f^{ap}$:
\begin{equation}
-\partial_{\tau} f^{ap} + \frac{1}{2} \sigma^2 r^{2 \gamma} \left[ \left(\partial _{r} f^{ap} \right)^2 + \partial^2_{r} f^{ap} \right] + (\alpha + \beta r) \partial_r f^{ap} - r = h(\tau,r).
\label{eq-lnP_aprox}
\end{equation}
If we insert the approximate solution into (\ref{PDE-1f}) then, after long but straightforward calculations based on expansion of all terms into a Taylor series in $\tau$ we obtain:
%\footnote{We employed software Mathematica to perform all symbolic calculations.}
\begin{equation}
h(\tau,r) = k_4(r) \tau^4 + k_5(r) \tau^5 + o(\tau^5)
\label{h}
\end{equation}
where $k_4$ and $k_5$  are given by 
\begin{eqnarray}
\label{k_4(r)}
k_4(r) &=& \frac{1}{24} \gamma r^{2(\gamma-2)} \sigma^2 
\left[ 
2 \alpha^2 (-1+2 \gamma) r^2 + 4 \beta^2 \gamma r^4 
- 8 r^{3+2\gamma} \sigma^2 
\right.
\nonumber \\ 
& &
+ 2 \beta (1-5 \gamma + 6 \gamma^2) r^{2 (1+\gamma)} \sigma^2 
+\sigma^4 r^{4 \gamma} (-3+16 \gamma - 28 \gamma^2 + 16 \gamma^3)
 \nonumber \\
& &
\left.
+ 2 \alpha r \left( \beta(-1+4 \gamma)r^2 + (2-7 \gamma + 6 \gamma^2) r^{2 \gamma} \sigma^2 \right)
 \right],
\end{eqnarray}
\begin{eqnarray}
\label{k5}
k_5(r)&=&
\frac{\gamma\sigma^2}{120}
 r^{2 \left( -2 + {\gamma} \right) }  
    \left[ 6 {\alpha}^2 \beta \left( -1 + 2 {\gamma} \right)  r^2 + 12 {\beta}^3 {\gamma} r^4 
 - 10 {\left( 1 - 2 {\gamma} \right) }^2 r^{1 + 4 {\gamma}} {\sigma}^4
\right. \nonumber \\
&&
+ 6 {\beta}^2  \sigma^2\left( 1 - 5 {\gamma} + 6 \gamma^2 \right)  r^{2 \left( 1 + {\gamma} \right) } 
 \nonumber \\
&&
 + \beta r^{2 \gamma}\sigma^2 \left( -10 \left( 5 + 2 \gamma \right)  r^3 +  
         3 {\left( 1 - 2 {\gamma} \right) }^2 \left( -3 + 4 {\gamma} \right)  r^{2 {\gamma}} {\sigma}^2 \right)  \nonumber \\
&& 
+ 2 \alpha r \biggl( 3 {\beta}^2 \left( -1 + 4 {\gamma} \right)  r^2 + 
         3 \beta \left( 2 - 7 {\gamma} + 6 {{\gamma}}^2 \right)  r^{2 {\gamma}} {\sigma}^2 
\nonumber \\
&& 
\qquad - \left. 5 \left( -1 + 2 {\gamma} \right)  r^{1 + 2 {\gamma}} {\sigma}^2 \biggr)  \right]\,.
\end{eqnarray}

Let us consider a function $g(\tau,r)= f^{ap} - f^{ex}$. As  
$\left( \partial_{r }g \right)^2 
=\left(\partial_{r} f^{ap} \right)^2  -\left(\partial_{r} f^{ex} \right)^2  -
2 \partial_{r} f^{ex} \partial_{r} g$ we have
\begin{eqnarray*}
-\partial_{\tau} g &+& \frac{1}{2} \sigma^2 r^{2 \gamma} 
\left[ 
\left(\partial_{r} g \right)^2 + \left(\partial^2_{r} g\right)
\right] +
(\alpha+\beta r) \partial_{r} g 
\\
&=& 
\left\{ 
-\partial_{\tau} f^{ap} + \frac{1}{2} \sigma^2 r^{2 \gamma} 
\left[ 
\left(\partial_{r} f^{ap} \right)^2 + \partial^2_{r} f^{ap}
\right] +
(\alpha+\beta r)\partial_{r} f^{ap} 
\right\} 
\\ 
& & -
\left\{ 
-\partial_{\tau}f^{ex} + \frac{1}{2} \sigma^2 r^{2 \gamma} 
\left[ 
\left(\partial_{r} f^{ex} \right)^2 + \left(\partial^2_{r} f^{ex} \right)
\right] +
(\alpha+\beta r) \partial_{r} f^{ex}
\right\}
 \\ 
& & -
\sigma^2 r^{2 \gamma} \partial_{r} f^{ex} \partial_{r}g\,.
\end{eqnarray*}
It follows from (\ref{eq-lnP}) and (\ref{eq-lnP_aprox}) that the function $g$ satisfies the following PDE:
we obtain a PDE for the function $g$:
\begin{eqnarray}
\label{pde-g}
-\partial_{\tau} g &+& \frac{1}{2} \sigma^2 r^{2 \gamma} 
\left[ \left( \partial_{r} g \right)^2 + \partial^2_{r} g 
\right] +(\alpha+\beta r) \partial_{r} g \nonumber \\
&=& h(\tau,r)-\sigma^2 r^{2 \gamma} (\partial_{r} f^{ex}) (\partial_{r} g),
\end{eqnarray}
where $h(\tau,r)$  satisfies (\ref{h}). Let us expand the solution of (\ref{pde-g}) into  a Taylor series  with respect to $\tau$ with coefficients depending on $r$. We obtain
$
g(\tau,r)=\sum_{i=0}^{\infty} c_i(r) \tau^i = \sum_{i=\omega}^{\infty} c_i(r) \tau^i,
$
i.e. the first nonzero term in the expansion is $c_{\omega}(r) \tau^{\omega}$.
Then
$
\partial_{\tau} g = \omega c_{\omega}(r) \tau^{\omega-1} + o(\tau^{\omega-1})
$ and
$
h(\tau,r)=k_4(r) \tau^{4} + o(\tau^4)
$ as 
$
\tau \rightarrow 0^+.
$
Here the term $k_4(r)$ is given by (\ref{k_4(r)}). The remaining terms in (\ref{h}) are of the order $o(\tau^{\omega-1})$ as $\tau \rightarrow 0^+$. Hence
$-\omega c_{\omega}(\tau) = k_4(r) \tau^4$
from which we deduce, for $\omega=5$, $c_{5}(r) = -\frac{1}{5} k_4(r)$.
It means that
$
g(\tau,r)=\ln P^{ap}(\tau,r) - \ln P^{ex}(\tau,r) = -\frac{1}{5} k_4(r) \tau^5 + o(\tau^5)
$
which completes the proof.
\hfill$\diamondsuit$
%------------------------------------------------------------

% --------------- corollaries ----------------------------
\begin{corollary}\label{cor:1}
Theorem \ref{theorem-accuracy-for-ln} enables us to compute error in yield curves which are given by 
 $R(\tau,r)=-\frac{\ln P(\tau,r)}{\tau}$ and relative error in bond prices.
\begin{enumerate}
\item The error in yield curves can be expressed as
\[R^{ap}(\tau,r)-R^{ex}(\tau,r)=-c_5(r) \tau^4 + o(\tau^4)  \hbox{ as } \tau \rightarrow 0^+;
\]
\item The relative error\footnote{This is referred to as the relative mispricing 
in \cite{1f-approximation}}  of $P$ is given by
\[
\frac{P^{ap}(\tau,r)-P^{ex}(\tau,r)}{P^{ex}(\tau,r)}=-c_5(r) \tau^5 + o(\tau^5) \hbox{ as } \tau \rightarrow 0^+.
\]
\end{enumerate}
The convergence is uniform w. r. to $r$ on compact subintervals $[r_1,r_2]  \subset\subset (0,\infty)$.
\end{corollary}
%------------------------------------------------------------

% -------proof of the corollaries -------------
{\em Proof:} 
The first corollary follows from the formula for calculating yield curves. To prove the second statement we note that Theorem \ref{theorem-accuracy-for-ln} gives $\ln P^{ap} - \ln P^{ex} = c_5(r) \tau^5 + o(\tau^5).$ Hence
$P^{ap}/P^{ex} = e^{c_5(r) \tau^5 + o(\tau^5)} = 1 + c_5(r) \tau^5 + o(\tau^5)$
and therefore
$\frac{P^{ap}-P^{ex}}{P^{ex}}=- c_5(r) \tau^5 + o(\tau^5).$
\hfill$\diamondsuit$
%------------------------------------------------------------

% --------------- remark - CIR model ----------------------------
\begin{remark}
For the CIR model with $\gamma=1/2$ the term $k_4(r)$ defined in  (\ref{k_4(r)}) can be simplified to
$\frac{1}{24} \sigma^2 \left[ \alpha \beta + r(\beta^2 - 4 \sigma^2) \right]$
and hence 
\[
\ln P^{ap}_{CIR}(\tau,r) - \ln P^{ex}_{CIR}(\tau,r) = -\frac{1}{120} \sigma^2 \left[ \alpha \beta + r(\beta^2 - 4 \sigma^2) \right] \tau^5 + o(\tau^5)
\]
as $\tau \rightarrow 0^+$ uniformly w. r. to $r$ on compact subintervals $[r_1,r_2]  \subset\subset [0,\infty)$.
\end{remark}
%------------------------------------------------------------

\subsection{Improved higher order approximation formula}
\label{higherorder}
It follows from (\ref{theorem-accuracy-for-ln})  that the term  
$\ln P^{ap}(\tau,r) - c_5(r) \tau^5$ is the higher order accurate approximation of $\ln P^{ex}$ when compared to  the original approximation 
$\ln P^{ap}(\tau,r)$ from  \cite{1f-approximation}. Furthemore, we show, that it is even possible to compute $O(\tau^6)$ term and to obtain a new approximation $\ln P^{ap2}(\tau,r)$ such that the difference $\ln P^{ap2}(\tau,r) - \ln P^{ex}(\tau,r)$ is $o(\tau^6)$ for small values of $\tau>0$.

Let $P^{ex}$ be the exact bond price in the model (\ref{1f-short-rate}). Let us define an improved approximation $P^{ap2}$ by the formula 
\begin{equation}
\ln P^{ap2}(\tau,r)  = \ln P^{ap}(\tau,r) - c_5(r) \tau^5 - c_6(r) \tau^6
\label{approx-higher-order}
\end{equation}
where $\ln P^{ap}$ is given by (\ref{1f-approximation-formula}), $c_5(\tau)$ is given by (\ref{c5(r)}) in Theorem 1 and
\[
c_6(r)=\frac{1}{6} \left( \frac{1}{2} \sigma^2 r^{2 \gamma} c_5''(r) + (\alpha+\beta r) c_5'(r) - k_5(r) \right)
\]
where $c_5^\prime$ and $c_5^{\prime\prime}$ stand for the first and second derivative of $c_5(r)$ w. r. to $r$ and $k_5$ is defined in (\ref{k5}).

% ------------- theorem --- label{theorem-accuracy-for-ln} ----------

\begin{theorem}\label{theorem-accuracy2-for-ln}
The difference between the higher order approximation $\ln P^{ap2}$ given by (\ref{approx-higher-order}) and the exact solution $\ln P^{ex}$ satisfies
$
\ln P^{ap2}(\tau,r) - \ln P^{ex}(\tau,r) = o(\tau^6)
$
as $\tau \rightarrow 0^+$. The convergence is uniform w. r. to $r$ on compact subintervals $[r_1,r_2]  \subset\subset (0,\infty)$.

\end{theorem}
%------------------------------------------------------------

% -------proof of the theorem -------------
{\em Proof:} 
We have to prove that $g(\tau,r)=c_5(r)\tau^5 + c_6(r) \tau^6 + o(\tau^6)$ where $c_5$ and $c_6$ are given above. We already know the form of the coefficient $c_5=c_5(r)$. Consider the following Taylor series expansions:
\[
g(\tau,r) = \sum_{i=5}^{\infty} c_i(r) \tau^i,\quad
h(\tau,r)=\sum_{i=4}^{\infty} k_i(r) \tau^i, \quad
f(\tau,r)=\sum_{i=1}^{\infty} l_i(r) \tau^i.
\]
The absolute term $l_0$ is zero because  $f^{ex}(0,r)=\ln P^{ex}(0,r) = \ln 1 =0$ for all $r>0$.
Substituting power series into equation (\ref{pde-g}) and comparing coefficients of the order $\tau^5$ enables us to derive the identity:
$
-6 c_6(r) + \frac{1}{2} \sigma^2 r^{2 \gamma} c_5''(r) + (\alpha+\beta r) c_5'(r) -k_5(r) =0
$
and hence
$
c_6(r)=\frac{1}{6} \left( \frac{1}{2} \sigma^2 r^{2 \gamma} c_5''(r) + (\alpha+\beta r) c_5'(r) - k_5(r) \right)
$
The term $k_5(r)$ given by (\ref{k5}) is obtained by computing the expansion of $h$. 
\hfill$\diamondsuit$

The order of relative error of bond prices and order of error of interest rates for the new higher order approximation can be derived similarly as in Corollary~\ref{cor:1}.
% --------------- remark - next terms + CIR model ----------------------------
\begin{remark}
It is not obvious how to obtain the next higher order terms of expansion  because the equations contain unknown coefficients $l_i(r)$, $i \geq 1$, of logarithm of the exact solution which is not known explicitly.
\end{remark}

\begin{remark}
In the case of the CIR model we have
\[
c_5^{CIR}(r)=-\frac{\sigma^2}{120}  \left( \alpha \beta + r(\beta^2 - 4 \sigma^2) \right), \: 
k_5^{CIR}(r)=\frac{ \beta \sigma^2}{40} \left( \alpha \beta +  (\beta^2 - 10 \sigma^2) r \right) 
\]
and so
$
c_6^{CIR}(r) = \frac{\sigma^2}{360}  \left( -2 \alpha \beta^2 + 17 \beta \sigma^2 r - 2 \beta^3 r +2 \alpha \sigma^2 \right).
$
Hence 
\begin{eqnarray*}
\ln P_{CIR}^{ap2} = \ln P_{CIR}^{ap} &+& \frac{ \sigma^2 }{120}\left( \alpha \beta + r(\beta^2 - 4 \sigma^2) \right) \tau^5 \\
&& - \frac{ \sigma^2}{360} \left( -2 \alpha \beta^2 + 17 \beta \sigma^2 r - 2 \beta^3 r +2 \alpha \sigma^2 \right) \tau^6
\end{eqnarray*}
The theorem yields $\ln P_{CIR}^{ap2}(\tau,r) - \ln P_{CIR}^{ex}(\tau,r) = o(\tau^6).$
By computing the expansions of both exact and this approximative solutions we finally obtain
\begin{eqnarray*}
\ln P_{CIR}^{ap2}(\tau,r)  &=& \ln P_{CIR}^{ex}(\tau,r)  
-\frac{\sigma^2}{5040}  \biggl( 11  {\alpha} {\beta}^3 + 
         11 {\beta}^4 r - 34  {\alpha} \beta {\sigma}^2 \\
&&- 180 {\beta}^2 r {\sigma}^2 + 34 r {\sigma}^4 \biggr) \tau^7 + o(\tau^7) \ \  \hbox{as}\ \  \tau \rightarrow 0^+.
\end{eqnarray*}
\end{remark}

\subsection{Comparison of approximations to the exact solution for the CIR model}
\label{exactsol}

In this section we present a comparison of the original and improved approximations in the case of the CIR model where the exact solution is known. We use the parameter values from 
\cite{1f-approximation}, i.e. $\alpha=0.00315$, $\beta=-0.0555$ and $\sigma=0.0894$.

In Table \ref{tab2}  we show $L_{\infty}$ and $L_2\,-\,$norms with respect to $r$ of the difference $\ln P^{ap} - \ln P^{ex}$ and $\ln P^{ap2} - \ln P^{ex}$ where we considered $r \in [0,0.15]$. Maximum value considered 0.15 means 15 percent interest rate, which should be sufficient for practical use. We also compute the experimental order of convergence (EOC) in these norms.
Recall that the experimental order of convergence gives an approximation of the exponent $\alpha$ of expected power law estimate for the error $\Vert\ln P^{ap}(\tau,.) - \ln P^{ex}(\tau,.)\Vert = O(\tau^{\alpha})$ as $\tau \rightarrow 0^+$. The $EOC_i$ is given by a ratio
\[
EOC_i = \frac{\ln (err_i/err_{i+1})}{\ln ( \tau_i/\tau_{i+1})}
\quad\hbox{where }\ \ err_i = \Vert\ln P^{ap}(\tau_i,.) - \ln P^{ex}(\tau_i,.)\Vert_p\,.
\]

\begin{table}
\caption{The $L_{\infty}$ and $L_2\,-\,$errors for the original $\ln P_{CIR}^{ap}$ and improved $\ln P_{CIR}^{ap2}$ approximations}
\begin{center}
\vglue 2mm
\scriptsize
\begin{tabular}{c|c|c|c|c} 
\hline
\hline
$\tau$ &  $\Vert\ln P^{ap} - \ln P^{ex}\Vert_{\infty}$ &  EOC &  $\Vert\ln P^{ap2} - \ln P^{ex}\Vert_{\infty}$ & EOC\\
\hline
1 &  $ 2.774 \times 10^{-7}$ &  4.930 &$4.682 \times 10^{-10}$  & 7.039 \\ 
0.75 &  $6.717 \times 10^{-8}$ &  4.951& $ 6.181 \times 10^{-11}$& 7.029 \\ 
0.5 &  $9.023 \times 10^{-9}$ &  4.972 &$3.576 \times 10^{-12}$& 7.004 \\ 
0.25 & $2.876 \times 10^{-10}$  &  --  &$2.786 \times 10^{-14}$& -- \\ 
\hline
\end{tabular}
\end{center}

\begin{center}
\vglue 2mm
\scriptsize
\begin{tabular}{c|c|c|c|c} 
\hline
\hline
$\tau$ &  $\Vert\ln P^{ap} - \ln P^{ex}\Vert_{2}$ &  EOC &  $\Vert\ln P^{ap2} - \ln P^{ex}\Vert_{2}$ & EOC\\
\hline
1 &  $6.345 \times 10^{-8}$ &  4.933 & $9.828 \times 10^{-11}$ & 7.042\\ 
0.75 & 1.535 $\times 10^{-8}$ &  4.953 & $1.296 \times 10^{-11}$ & 7.031\\ 
0.5 & 2.061 $\times 10^{-9}$ & 4.973 & $7.492 \times 10^{-13}$ & 7.012\\ 
0.25 &  6.563 $\times 10^{-11}$ &  --  & $5.805 \times 10^{-15}$ & -- \\
\hline
\end{tabular}
\end{center}

\label{tab2}
\end{table} 

\begin{figure}

\begin{center}
\includegraphics[width=7cm]{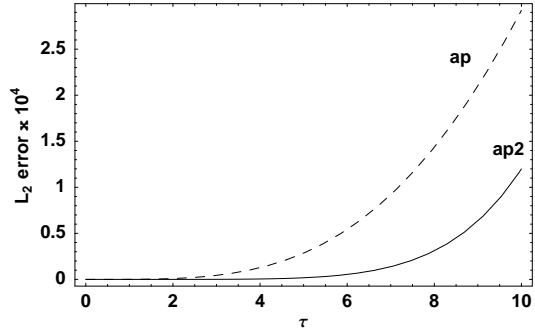}
\end{center}

\caption{The error $\Vert\ln P^{ap}(\tau, .) - \ln P^{ex}(\tau, .)\Vert_{2}$  for the original approximation (dashed line) and the new approximation (solid line). Horizontal axis is time to maturity $\tau$.}

\label{fig:1}
\end{figure}

In Table \ref{tab1} and Figure \ref{fig:1} we show the $L_2\,-\,$error of the difference between the original and improved approximations for larger values of $\tau$. It turned out that the higher order approximation $P^{ap2}$ gives about twice better approximation of bond prices in the long time horizon up to 10 years.

\begin{table}
\caption{The $L_2\,-\,$error with respect to $r$ for large values of $\tau$.}

\begin{center}
\vglue2mm
\scriptsize
\begin{tabular}{c|c|c|c|c|c} 
\hline
\hline
$\tau$ & 1 & 2 & 3 &  4  & 5  \\
\hline
$\Vert\ln P^{ap} - \ln P^{ex}\Vert_{2}$  &  
$6.345 \times 10^{-8}$ &  
$1.877 \times 10^{-6}$ &
$1.314 \times 10^{-5}$ &
$5.093 \times 10^{-5}$ &
$1.427 \times 10^{-4}$ \\
$\Vert\ln P^{ap2} - \ln P^{ex}\Vert_{2}$ & 
$9.828 \times 10^{-11}$ &
$1.314 \times 10^{-8}$ &
$2.329 \times 10^{-7}$ &
$1.799 \times 10^{-6}$ &
$8.798 \times 10^{-6}$ \\
\hline 
\end{tabular}
\end{center}

\begin{center}
\vglue3mm
\scriptsize
\begin{tabular}{c|c|c|c|c|c} 
\hline
\hline
$\tau$ & 6 & 7 & 8 &  9  & 10  \\
\hline
$\Vert\ln P^{ap} - \ln P^{ex}\Vert_{2}$  &  
$3.255 \times 10^{-4}$ &
$6.441 \times 10^{-4}$ &
$1.148 \times 10^{-3}$ &
$1.890 \times 10^{-3}$ &
$2.921 \times 10^{-3}$ \\ 
$\Vert\ln P^{ap2} - \ln P^{ex}\Vert_{2}$ & 
$3.217 \times 10^{-5}$ &
$9.618 \times 10^{-5}$ &
$2.479 \times 10^{-4}$ &
$5.705 \times 10^{-4}$ &
$1.200 \times 10^{-3}$ \\
\hline
\end{tabular}
\end{center}

\label{tab1}
\end{table} 

\subsection{Comparison of approximate and numerical solutions}
\label{numsol}

In Table \ref{tabnum1} we present a comparison of the  original approximation formula with a numerical solution $P^{num}$. The numerical solution was obtained using a finite volume method. 
We used $10^5$ spatial and $4.10^{7}$ time discretization grid points in the computational domain $\tau \in [0,1]$, $r \in [0,0.5]$ in order to achieve the $L_2\,-\,$errors less than $10^{-11}$ between exact solution for the CIR model and the numerical solution.
The difference $O(10^{-11})$ between the numerical and approximate solutions is therefore of the same order of accuracy as the numerical scheme and hence it was not reasonable to compute EOC in this case.

\begin{table}
\caption{Norms of the difference $\ln P^{ap}(\tau,.) - \ln P^{num}(\tau,.)$  for several values of $\tau$ and $\gamma$.}

\begin{center}
\vglue2mm
\scriptsize
\begin{tabular}{c|c|c|c|c} 
\hline
\hline
 ~& \multicolumn{2}{|p{110pt}}{\qquad\qquad $\gamma=0.5$} &  \multicolumn{2}{|p{110pt}}{\qquad\qquad\quad $\gamma=0.75$}  \\
\hline
$\tau$ & $L_{\infty}$ norm & $L_{2}$ norm  & $L_{\infty}$ norm & $L_{2}$ norm  \\
\hline
1 &  $2.771 \times 10^{-7}$&   $8.967 \times 10^{-8}$ &   $5.576 \times 10^{-8}$  &   $1.429 \times 10^{-8}$\\ 
0.75 &  $6.694 \times 10^{-8}$&  $2.165 \times 10^{-8}$ &   $1.691 \times 10^{-8}$ &   $3.429 \times 10^{-9}$ \\ 
0.5 &  $8.854 \times 10^{-9}$&  $2.867 \times 10^{-9}$ &   $1.411 \times 10^{-8}$ &   $4.656 \times 10^{-10}$ \\ 
0.25 &  $3.400\times 10^{-10}$&   $7.236 \times 10^{-11}$ &   $6.963 \times 10^{-9}$ &   $9.542 \times 10^{-11}$ \\
\hline 
\hline 
\end{tabular}
\end{center}

\begin{center}
\vglue3mm
\scriptsize
\begin{tabular}{c|c|c|c|c} 
\hline
\hline
 ~& \multicolumn{2}{|p{110pt}}{\qquad\qquad $\gamma=1.00$} &  \multicolumn{2}{|p{110pt}}{\qquad\qquad\quad $\gamma=1.32$}  \\
\hline
$\tau$ & $L_{\infty}$ norm & $L_{2}$ norm  & $L_{\infty}$ norm & $L_{2}$ norm  \\
\hline
1 &  $5.798 \times 10^{-9}$&   $1.296 \times 10^{-9}$ &   $2.664 \times 10^{-9}$  &   $5.536 \times 10^{-10}$\\ 
0.75 &  $1.216 \times 10^{-9}$&  $2.838 \times 10^{-10}$ &   $1.406 \times 10^{-9}$ &   $2.352 \times 10^{-10}$ \\ 
0.5 &  $9.071 \times 10^{-10}$&  $7.488 \times 10^{-11}$ &   $1.113 \times 10^{-9}$ &   $1.413 \times 10^{-10}$ \\ 
0.25 &  $6.154\times 10^{-10}$&   $5.663 \times 10^{-11}$ &   $7.860 \times 10^{-10}$ &   $8.524 \times 10^{-11}$ \\
\hline 
\end{tabular}
\end{center}
\label{tabnum1}
\end{table} 

\section{Conclusions}\label{concl}
We analyzed qualitative properties of the approximation formula for pricing zero coupon bonds due to 
Choi and Wirjanto \cite{1f-approximation}.  We furthermore proposed a higher order approximation formula for pricing zero coupon bonds. We derived the order accuracy for both approximations and we test them numerically. The improved approximation is more accurate for a reasonable range of time horizons.

\section*{Acknowledgments}
The authors thank the referees for their valuable comments. The support from grants DAAD-MSSR-11/2006, VEGA 1/3767/06 and UK/381/2007 is acknowledged.

%------------------------------------------------------------
\appendix
\section{Uniqueness of a solution to zero coupon bond PDE}
\label{uniqueness}
In this section, we give a proof of Theorem 1. Our aim is to prove the inequality 
\begin{equation}
\frac{\dd}{\dd\tau} \int_0^{\infty} r^{\omega} P^2 \dr \leq K \int_0^{\infty} r^{\omega} P^2 \dr
\label{ineq-uniqueness}
\end{equation}
to be satisfied by any solution of (\ref{PDE-1f}) with some constants $K$ and $\omega \geq 0$. It implies the uniqueness of a solution to the PDE (\ref{PDE-1f}). Indeed, if $P_1$ and $P_2$ are two solutions of (\ref{PDE-1f}) with the same initial condition $P(0,r)=1$. Then $P=P_1-P_2$ is also a solution to (\ref{PDE-1f}) with $P(0,r)=0$. Let us define a function 
$y(\tau)=\int_0^{\infty} r^{\omega} P^2(\tau,r) \dr$.
Then the inequality (\ref{ineq-uniqueness}) means 
$\frac{dy(\tau)}{d \tau} \leq K y(\tau)$ for $\tau >0$.
It implies:
$
\frac{d}{d \tau} \left( e^{-K \tau} y(\tau) \right) = -K e^{-K \tau} y(\tau) + e^{-K \tau}
 \frac{d y(\tau)}{d \tau} \leq 0.
$
Since $y(0)=0$ and $y(\tau) \geq 0$, it follows that $y(\tau)=0$ for all $\tau$. Thereof $P(\tau,r)=0$ for all $\tau \geq 0$, $r \geq 0$ and hence $P_1 \equiv P_2$ as claimed.

Now let us derive inequality (\ref{ineq-uniqueness}).
Multiplying the equation by $r^{\omega} P$, where $\omega > 0$ and $2 \gamma + \omega -1 >0$ using the identity
$ \frac{1}{2}\frac{d}{d \tau} \int_0^{\infty} r^{\omega} P^2 \dr= \int_0^{\infty} r^{\omega} P \partial_{\tau} P \dr$, and integrating with respect to $r$ from 0 to infinity we obtain\footnote{In what follows, we shall omit the differential $\hbox{d}r$ from the notation}
\begin{equation}
\hbox{\hglue-5truemm}
\frac{1}{2}\frac{\hbox{d}}{\hbox{d}\tau} \int_0^{\infty} \hskip -2truemm r^{\omega} P^2
= \frac{\sigma^2}{2} \int_0^{\infty} \hskip -2truemm r^{2 \gamma + \omega} \partial^2_{r} P P   +
\int_0^{\infty} \hskip -2truemm (\alpha+\beta r) r^{\omega}  \partial_{r} P P   - \int_0^{\infty} \hskip -2truemm r^{\omega+1} P^2.
\label{eq-1}
\end{equation}

We use the notation $P'=\partial_{r} P$, $P''=\partial^2_{r} P$. 
Firstly, we use integration by parts for the following integrals from the above equation:
\begin{eqnarray*}
\int_0^{\infty} r^{2 \gamma +\omega} P'' P   &=& -(2 \gamma+\omega) \int_0^{\infty} r^{2 \gamma +\omega-1} P P'   - \int_0^{\infty} r^{2 \gamma  + \omega} (P')^2  \\
&=&\frac{1}{2}(2 \gamma+\omega)(2 \gamma+\omega-1) \int_0^{\infty} r^{2 \gamma+\omega-2} P^2   - \int_0^{\infty} r^{2 \gamma  + \omega} (P')^2
\end{eqnarray*}
where we have used the identity $\int_0^{\infty}r^{\omega+\xi} P'P   = - \frac{\omega+\xi}{2} \int_0^{\infty}r^{\omega+\xi-1} P^2$ valid for any $\omega,\xi\ge 0$ and a function $P$ satisfying the decay estimates from Definition~\ref{def:1}. Substituting this to (\ref{eq-1}), we end up with the identity 
\begin{eqnarray}
\label{eq-2}
\frac{1}{2} \frac{\dd}{\dd\tau} \int_0^{\infty} r^{\omega} P^2   &=&
\frac{\sigma^2}{4}(2 \gamma+\omega)(2 \gamma+\omega-1) \int_0^{\infty} r^{2 \gamma+\omega-2} P^2   - \frac{\sigma^2}{2} \int_0^{\infty} r^{2 \gamma  + \omega} (P')^2  
\nonumber \\
&-& \frac{\alpha \omega}{2} \int_0^{\infty}  r^{\omega-1} P^2   - \frac{(\omega+1) \beta}{2} \int_0^{\infty}  r^{\omega} P^2   - \int_0^{\infty}  r^{\omega+1} P^2.
\end{eqnarray}

\textbf{Case 1: $\gamma = \frac{1}{2}$ and $2 \alpha \geq \sigma^2$.} 
We recall that the condition $2 \alpha \geq \sigma^2$ in the case of CIR model ($\gamma=\frac12$) is very well understood as it almost surely guarantees the strict positivity of the stochastic processes $r=r_t$ satisfying the stochastic differential equation: 
$\dr= \left( \alpha+\beta r \right) \dt + \sigma \sqrt{r} \dw$ (see e.g. \cite{K}).

\textbf{Subcase 1a: $2 \alpha > \sigma^2$.}
We use the equality (\ref{eq-2}) with $\gamma=1/2$ and $\omega=\frac{2 \alpha}{\sigma^2} - 1>0$ to obtain the desired inequality
(\ref{ineq-uniqueness}) with $K=(\omega+1)\beta$.
%\[
%\frac{1}{2} \frac{\hbox{d}}{\hbox{d}\tau} \int_0^{\infty} r^{\omega} P^2 \leq  -\frac{(\omega+1) \beta}{2} \int_0^{\infty}  r^{\omega} P^2.
%\]

\par
\textbf{Subcase 1b: $2 \alpha = \sigma^2$.} 
Using identity (\ref{eq-2}) with $\omega=0$ (or simply by multiplying the PDE with $P$ and integrating over $(0,\infty)$) we obtain the inequality
(\ref{ineq-uniqueness}) with $K=\beta$.

\par
\textbf{Case 2: $\gamma \in \left(\frac{1}{2}, 1 \right)$.}
We use equation (\ref{eq-1}) with $\omega=2$ and estimate the integral $\int_0^{\infty} r^{2 \gamma} P^2$  by using H\"older's inequality:
\[
\int_0^{\infty} r^{2 \gamma} P^2 = \int_0^{\infty} \left( r^{4\gamma-2} P^{4\gamma-2} \right) \left(r^{2-2\gamma} P^{4-4\gamma)} \right)   \leq \left( \int_0^{\infty} r^2 P^2   \right)^{2 \gamma-1} 
\left(\int_0^{\infty}r P^2   \right)^{2-2\gamma}.
\]
It follows from the Young's inequality $ab\le \frac{1}{p\varepsilon^p} a^p + \frac1q \varepsilon^q b^q$ for $p,q\ge1$ such that $\frac1p+\frac1q=1$ and any $\varepsilon >0$ we get
\[
\int_0^{\infty} r^{2 \gamma} P^2
\leq 
(2\gamma-1) \left(\frac{1}{\varepsilon} \right)^{\frac{1}{2 \gamma-1}} \int_0^{\infty} r^2 P^2   +
(2-2\gamma) \varepsilon^{\frac{1}{2 \gamma-2}} \int_0^{\infty} r P^2 . 
\]
Again using (\ref{eq-2}) with $\omega=2$ and the above estimate we obtain
\begin{eqnarray*}
\frac{1}{2} \frac{\dd}{\dd\tau} \int_0^{\infty} r^2 P^2   
&\leq& 
\frac{\sigma^2}{2} (\gamma+ 1)(2 \gamma+1) \int_0^\infty r^{2 \gamma} P^2   -  \alpha \int_0^{\infty}  r P^2   - 
 \frac{3 \beta}{2} \int_0^{\infty}  r^2 P^2  \\
&\leq& K \int_0^{\infty} r^2 P^2    + 
\left(\sigma^2 (\gamma+ 1)(2 \gamma+1)  (1-\gamma) \varepsilon^{\frac{1}{2-2 \gamma}} -\alpha \right) \int_0^{\infty} r P^2. 
\end{eqnarray*}
where $K=\frac{\sigma^2}{2} (\gamma+ 1)(2 \gamma+1)(2\gamma-1) \left(\frac{1}{\varepsilon} \right)^{\frac{1}{2 \gamma-1}} - \frac{3 \beta}{2}$. By choosing $\varepsilon>0$ sufficiently small such that 
$\sigma^2 (\gamma+ 1)(2 \gamma+1) (1-\gamma) \varepsilon^{\frac{1}{2-2 \gamma}} -\alpha < 0$, we finally obtain the desired inequality $\frac{1}{2} \frac{d}{d\tau} \int_0^{\infty} r^2 P^2 \le K \int_0^{\infty} r^2 P^2$.

\par
\textbf{Case 3: $\gamma = 1$.}
We again use the equation (\ref{eq-2}) with $\omega=2$. we obtain 
(\ref{ineq-uniqueness}) with $K=3 (2 \sigma^2 - \beta)$.

\par
	\textbf{Case 4: $\gamma \in \left(1, \frac{3}{2} \right)$.} 
Similarly as in the case $\frac12<\gamma<1$ we make use of the H\"older's inequality integral estimation:
\[
\hbox{\hglue-8truemm}
\int_0^{\infty} r^{2 \gamma} P^2 = \int_0^{\infty} \left( r^{6-4\gamma} P^{6-4\gamma} \right) \left(r^{6\gamma-6} P^{4\gamma-4} \right)   
\leq \left( \int_0^{\infty} r^2 P^2   \right)^{3-2 \gamma} \left(\int_0^{\infty}r^3 P^2   \right)^{2\gamma-2}
\]
and, by Young's inequality, we obtain, for any $\varepsilon>0,$
\[
\int_0^{\infty} r^{2 \gamma} P^2\leq
(3-2\gamma) \left(\frac{1}{\varepsilon} \right)^{\frac{1}{3-2 \gamma}}\int_0^{\infty} r^2 P^2  +
(2\gamma-2) \varepsilon^{\frac{1}{2 \gamma-2}} \int_0^{\infty} r^3 P^2 .
\]
By (\ref{eq-2}) with $\omega=2$ we have
\begin{eqnarray*}
\frac{1}{2} \frac{\hbox{d}}{\hbox{d}\tau} \int_0^{\infty} r^2 P^2   
&\leq& \frac{\sigma^2}{2} (\gamma+ 1)(2 \gamma+1) \int_0^\infty r^{2 \gamma} P^2  - \frac{3 \beta}{2} \int_0^{\infty}  r^2 P^2  - \int_0^{\infty}  r^3 P^2 \\
&\leq& K\int_0^{\infty} r^2 P^2   
+ \left(\sigma^2 (\gamma+ 1)(2 \gamma+1)  (\gamma-1) \varepsilon^{\frac{1}{2 \gamma-2}} -1 \right) \int_0^{\infty} r^3 P^2. 
\end{eqnarray*}
where $K= \frac{\sigma^2}{2} (\gamma+ 1)(2 \gamma+1)(3-2\gamma) \left(\frac{1}{\varepsilon} \right)^{\frac{1}{3-2 \gamma}} - \frac{3 \beta}{2}$. By choosing $\varepsilon>0$ sufficiently small such that  
$\sigma^2 (\gamma+ 1)(2 \gamma+1)  (\gamma-1) \varepsilon^{\frac{1}{2 \gamma-2}} -1  <0$ we end up with the desired inequality
$\frac{1}{2} \frac{d}{d \tau} \int_0^{\infty} r^2 P^2  \le K\int_0^{\infty} r^2 P^2$.


\begin{thebibliography}{}


\bibitem{1f-approximation}
Y. Choi and T. Wirjanto,
An analytic approximation formula for pricing zero-coupon bonds,
Finance Research Letters {\bf 4} (2007), 
116--126.


\bibitem{CKLS} 
K.L. Chan, G.A. Karolyi, F.A. Longstaff, and A.B. Sanders,
An Empirical Comparison of Alternative Models of the Short-Term Interest Rate, 
Journal of Finance {\bf 47} (1992), 
1209--1227.


\bibitem{CIR} 
J. Cox, K. Ingersoll and S. Ross, 
A Theory of the Term Structure of Interest Rates, 
Econometrica {\bf 53} (1985)
385--407.

\bibitem{K} 
Y.K. Kwok, 
Mathematical Models of Financial Derivatives,
New York, Heidelberg, Berlin: Springer Verlag, 1998.

\bibitem{N} 
K.B. Nowman,
Gaussian Estimation of Single-Factor Continuous Time Models 
of the Term Structure of Interest Rates,
Journal of Finance {\bf 52} (1997),
1695--1706.

\bibitem{TG}
S. Treepongkaruna and S. Gray,
On the Robustness of Short Term Interest Rate Models, 
Accounting and Finance {\bf 43} (2003), 
87--121.

\bibitem{V} 
O.A. Vasicek, 
An Equilibrium Characterization of the Term Structure,  
Journal of Financial Economics {\bf 5} (1977),
177--188.


\end{thebibliography}
\end{document}